\documentclass[12pt,preprint]{aastex}

\newcommand{\pifi}{\left[{\cal P}_i\right]\in{\cal F}_i\left[{\cal P}_i\right]}
\newcommand{\ponefone}{\left[{\cal P}_1\right]\in{\cal F}_1\left[{\cal P}_1\right]}
\newcommand{\ptwoftwo}{\left[{\cal P}_2\right]\in{\cal F}_2\left[{\cal P}_2\right]}
\newcommand{\pfourffour}{\left[{\cal P}_4\right]\in{\cal F}_4\left[{\cal P}_4\right]}
\newcommand{\pthfth}{\left[{\cal P}_3\right]\in{\cal F}_3\left[{\cal P}_3\right]}
\newcommand{\pgrcgr}{\left[{\cal P}_{GR}\right]\in{\cal C}_{GR}\left[{\cal P}_{GR}\right]}
\newcommand{\pgrdgr}{\left[{\cal P}_{GR}\right]\in{\cal D}_{GR}\left[{\cal P}_{GR}\right]}
\newcommand{\pgrfgr}{\left[{\cal P}_{GR}\right]\in{\cal F}_{GR}\left[{\cal P}_{GR}\right]}
\newcommand{\pgrggr}{\left[{\cal P}_{GR}\right]\in{\cal G}_{GR}\left[{\cal P}_{GR}\right]}

\shorttitle{Shocks in Black Hole Accretion}
\shortauthors{Das}

\begin{document}


\title{Generalized Shock Solutions for Hydrodynamic \\
Black Hole Accretion}


\author{Tapas K. Das}
\altaffiltext{1}{Present Address: Division of Astronomy, University of 
California at Los
Angeles, Box 951562, Los Angeles, CA 90095-1562, USA;
tapas@astro.ucla.edu}
\affil{Racah Institute of Physics, The Hebrew University of Jerusalem, Israel\\
IUCAA, Post Bag 4, Ganeshkhind, Pune
411 007, India tapas@iucaa.ernet.in}



\begin{abstract}
\noindent
For the first time, {\it all} available pseudo-Schwarzschild 
potentials are exhaustively used to investigate the possibility of
shock formation in hydrodynamic, invicid, black hole accretion discs.
It is shown that a significant region of parameter space spanned by
important accretion parameters allows shock formation
for flow in {\it  all} potentials used in this work.  This leads
to the conclusion that the standing shocks are essential 
ingredients in accretion discs around non-rotating black holes
in general.
Using a complete general relativistic framework, equations governing 
multi-transonic black hole accretion and wind are also formulated 
and solved in the Schwarzschild metric. Shock solutions for 
accretion flow in various pseudo potentials are then compared with such 
general relativistic solutions to identify which potential is
the best approximation of Schwarzschild space-time as far as the 
question of shock formation in black hole accretion discs is
concerned.
\end{abstract}


\noindent
\hrule
\begin{center}
{\bf Published in the The Astrophysical Journal, 2002, Volume 577, Issue 2, pp. 880-892.}
\end{center}
\hrule


\section{Introduction}
\noindent
The process by which any gravitating, massive,
astrophysical object captures its surrounding fluid
is called accretion. Depending on the rotational energy
content of the infalling material,
accretion flows onto  black holes may be broadly classified
into two different categories, i.e., 
non-rotating (spherical)  and rotating accretion (accretion
discs). 
If the instantaneous dynamical velocity and local acoustic velocity
 of the accreting fluid, moving along a space curve parameterized by $r$, are
$u(r)$ and $a(r)$ respectively, then the local Mach number $M(r)$ of the
 fluid can be defined as $M(r)=\frac{u(r)}{a(r)}$.
The flow will be locally
 subsonic or supersonic according to $M(r) < 1$ or $ >1$, i.e., according to
 $u(r)<a(r)$ or $u(r)>a(r)$. The flow is transonic if at any moment
 it crosses $M=1$. This happens when a subsonic to supersonic or supersonic to
 subsonic transition takes place either continuously or discontinuously.
The point(s) where such crossing 
takes place continiously is (are) called sonic point(s),
 and where such crossing takes place discontinuously are called shocks
 or discontinuities.
It is generally argued that, in order to
satisfy the inner boundary conditions imposed by the
event horizon, accretion onto black holes 
exhibit transonic properties in general, which further 
indicates that formation of shock waves
are  possible in astrophysical fluid flows onto
galactic and extra-galactic black holes. One also
expects that shock formation in black hole accretion
might be a general phenomena because shock waves
in rotating and non-rotating flows are convincingly
able to provide an important and efficient mechanism
for conversion of significant amount of the
gravitational energy (available from deep potential
wells created by these massive compact accretors) into
radiation by randomizing the directed infall motion of
the accreting fluid. Hence shocks possibly play an
important role in governing the overall dynamical and
radiative processes taking place in accreting plasma.
Thus the study of steady, stationary shock waves produced in black
hole accretion has acquired a very important status in recent
years and it is expected that shocks may be an
important ingredient in an accreting black hole
system in general. \\
\noindent
While the possibility of the formation of a standing
spherical shock around compact objects was first
conceived long ago (Bisnovatyi-Kogan,
Zel`Dovich, \& Sunyaev 1971), most of the works on
shock formation in spherical accretion share more or
less the same philosophy that one should incorporate
shock formation to increase the efficiency of directed
radial infall in order to explain the high luminosity
of AGNs and QSOs and to model their broad band
spectrum (Jones \& Ellison 1991).
Considerable work has been done in this direction
where several authors have investigated the formation
and dynamics of standing shock in spherical accretion
(M\'esz\'aros \& Ostriker 1983, Protheros \& Kazanas 1983,
Chang \& Osttriker 1985, Kazanas \& Ellision 1986,
Babul, Ostriker \& M\'esz\'aros 1989, Park 1990, 1990a).
Ideas and formalisms developed
in these works have been applied to study related
interesting problems like 
entropic-acoustic or various
other instabilities in spherical accretion (Foglizzo \& Tagger 2000,
Blondin \& Ellison 2001, Lai \& Goldreich 2000, Foglizzo 2001,
Kovalenko \& Eremin 1998),
production of high energy cosmic rays from AGNs
(Protheroe \& Szabo  1992), study of the hadronic model of AGNs
(Blondin  \& Konigl 1987, Contopoulos \& Kazanas 1995),  
high energetic emission from relativistic
particles in our galactic centre (Markoff, Melia \& Sarcevic 1999),
explanation of high lithium abundances in the
late-type, low-mass companions of the soft X-ray
transient, (Guessoum \& Kazanas 1999), study of accretion powered
spherical winds emanating from galactic and extra
galactic black hole environments (Das 2001).\\
\noindent
With equal (if not more) importance and rigor, the
question of shock formation in
accretion discs around Schwarzchild black holes has
been addressed by several authors. While the initial
works in this direction can be attributed to
Fukue (1983),  Hawley, Wilson \& Smarr (1984), Ferrari et al.
(1985),  Swada et al.  (1986) and Spruit (1987), it was Fukue (1987) and 
Chakrabarti and his
collaborators (Chakrabarti 1989 (C89 hereafter), 1996 and references therein,
Abramowicz \& Chakrabarti 1990,
Chakrabarti \& Molteni 1993) who were the first to provide
the satisfactory semi-analytical or numerical
global shock solution for transonic, invicid,
Keplerian or sub-Keplerian rotating accretion around a
Schwarzchild black hole. Consequently, their works
were further supported and improved by several other independent
works (Yang \& Kafatos 1995 (YK hereafter),
Caditz \& Tsuruta 1998, T\'oth, Keppens
\& Botchev 1998).
Because of the inner boundary conditions imposed by the event horizon,
shocks form in BH accretion discs only if the flow has more than one
real physical $X$ type
sonic point (multi-transonic flow).
For a particular set of initial
boundary conditions, some of the above mentioned works report
multiplicity in shock location, but such a degeneracy can ultimately be
removed by local stability analysis, 
allowing one to assert that only one stable shock location is possible.
 Hereafter, whenever we will use the word
`shock', it is to be understood that we will, in general,
always refer only the stable shock location
unless otherwise mentioned.\\
\noindent
The above mentioned works 
deserve attention because the shocked flows studied there are 
expected to explain 
the spectral properties of BH candidates.
However, thus far in the astrophysical 
literature, 
the theoretical study of steady, standing 
shock formation in accretion discs around non-rotating BHs
has suffered from two general limitations.
Firstly, the shock solutions were obtained either on a case
by case basis, or, even when successful attempts were made  to provide 
a more complete analysis,
the boundary of the parameter space
responsible for shock formation was obtained only for global variation of 
the total specific energy ${\cal E}$ (or accretion rate ${\dot M}$)
and specific angular momentum $\lambda$ of the flow, and not for
variations of
the polytropic constant $\gamma$ of the
flow; rather accretion was always considered to be ultra-relativistic
\footnote{By the term `ultra-relativistic' and `purely non-relativistic'
we mean a flow with
$\gamma=\frac{4}{3}$ and $\gamma=\frac{5}{3}$ respectively,
according to the terminology used in
Frank et. al. 1992.} 
which may {\it not} always be a realistic assumption.
As $\gamma$ is expected to 
have great influence on the radiative properties of the flow in general, we
think that ignoring the explicit 
dependence of shock solutions on $\gamma$ limits claims of generality. 
Secondly, except for YK,
all available so called global shock solutions have been discussed 
only in the context of one particular type of BH potential, namely,
the Paczy\'nski \& Wiita (1980) potential 
($\Phi_1$ hereafter).
Along with the $\Phi_1$,
recent studies (Das \& Sarkar 2001; hereafter DS, and references therein)
enhance the importance of also considering
three other pseudo-Schwarzschild BH potentials, one ($\Phi_2$ hereafter)
proposed by Nowak \& Wagoner (1991),
and two others ($\Phi_3$ and $\Phi_4$ hereafter)
due to Artemova, Bj\"{o}rnsson \& Novikov (Artemova, Bj\"{o}rnsson \& 
Novikov 1996, ABN hereafter), 
in mimicking the 
complete general relativistic space-time for accretion 
around a Schwarzschild black
hole.
Hence we believe that being restricted to only one specific 
pseudo-Schwarzschild BH potential 
does  not guarantee the claimed `global' nature of so called
global shock solutions present in the literature, rather one 
 must study the transonic disc structure as well as shock 
formation in {\it all} 
available BH potentials to firmly assert the ubiquity of shock 
formation in multi-transonic accretion disc around a Schwarzschild BH.
In this context, it is to be mentioned here that YK deserves special 
importance because the shock solution due to YK appears to be the 
only work available in the literature which provides the complete general 
relativistic description of shock formation {\it exclusively} 
for a non-rotating BH. Nevertheless, this work deals only 
with isothermal accretion 
but one understands that global
isothermality in BH accretion is difficult to achieve for realistic
flows and more general kind of BH accretion is expected to be 
governed by polytropic equation of state.
Also YK 
does not provide the global parameter space dependence of shock solutions.
A few authors claim to provide the full general relativistic 
shock solutions for Schwarzschild BHs 
as a limiting case of their results obtained in Kerr geometry (Chakrabarti
1996a,b; Lu, Yu, Yuan \& Young 1997, LY3 hereafter).
In doing so, a number of assumptions were made, some of which, however,
may not appear to be fully convincing. For example, either the disc
was supposed to be in conical equilibrium (LY3),
which should not be the case in
reality because the realistic accretion flow should be in vertical 
equilibrium (Chakrabarti 1996 and references therein); 
some results valid for isothermal accretion were directly
applied to study the polytropic accretion in an ad-hoc manner (Chakrabarti 
1996a), or some of the Newtonian approximations were
not very convincingly combined with complete general relativistic equations 
(Chakrabarti 1996b) which does not strengthen their
claim for a full general relativistic treatment of shock formation.
Hence it is fair to say that although literature on general relativistic
hydrodynamic BH accretion is well enriched by a number of important 
works (The following is an incomplete list of relevant papers on the subject:
Novikov  \& Thorne  1973, NT hereafter, Bardeen  \& Petterson 1975,
Abramowicz, Jaroszynski \& Sikora 1978, Lu  1985,1986, Karas
\& Mucha 1993, Bj\"{o}rnsson 1995, Riffert  \& Herold  1995,
Ipser  1996, Pariev  1996, Peitz \& Appl  1997,  Bao, Wiita \& Hadrava  1998,
Gammie \& Popham  1998,  Gammie 1999),
no well accepted complete
general relativistic global shock solution exclusively obtained
for a hydrodynamic accretion disc
around a Schwarzschild BH has yet appeared in the literature.\\
Motivated by above mentioned limitations encountered by
previous works in this field, the major aim of our
work presented in this paper is to provide a
generalized formalism which is expected to handle the
formation of steady, standing Rankine- Hugoniot
shock (RHS) in multi-transonic hydrodynamic BH accretion
flow and to identify which region of parameter space
(spanned by every important accretion parameter,
namely, ${\cal E}$, $\lambda$ and
$\gamma$), will be responsible for such shock
formation for {\it all} available pseudo- Schawarzschild BH
potentials. We would also like to
compare the properties of multi transonic accretion in
these BH potentials with complete general
relativistic BH accretion as long as the issue of shock
formation is concerned.\\
Hereafter, 
 we will define the Schwarzschild radius $r_g$ as
$$
r_g=\frac{2G{M_{BH}}}{c^2}
$$
(where  $M_{BH}$  is the mass of the black hole, $G$
is universal gravitational
constant) so that the marginally bound
circular orbit $r_b$ and the last stable circular orbit $r_s$
take the values $2r_g$
and $3r_g$ respectively for a typical Schwarzschild black hole. Also,
total
mechanical energy per unit mass on $r_s$ (sometimes called
`efficiency' $e$) may be computed as $-0.057$ for this
case. Also, we will use a simplified geometric unit throughout this paper where
radial distance $r$ is scaled in units of $r_g$, radial dynamical
velocity
$u$ and polytropic sound speed $a$ of
the flow is scaled in units of $c$ (the
velocity
of light in vacuum), mass $m$ is scaled in units of $M_{BH}$
and all other derived quantities would be scaled
accordingly. Also, for simplicity,we will use $G=c=1$.
In next section, we will briefly describe a few
important features of the four different pseudo-Schwarzschild 
`effective' BH
potentials used in this work. In \S 3, we will show
how we formulate and solve the equations governing
multi-transonic BH accretion in these potentials which
may have shocks. In \S 4, we will study
multi-transonic BH accretion using the full general
relativistic frame work and will try to argue (in an
indirect, but self consistant manner) which potential
is expected to be the closest approximation of actual
general relativistic solutions for which regions of
parameter space spanned by 
${\cal E}$, $\lambda$ and
$\gamma$,
as long as one
concentrates only on shocked flows. Finally in \S 5
we will draw our conclusion by highlighting some of the possible
important
impacts of study of shock formation on related
fields.
\section {Properties of four pseudo- Schwarzschild BH
potentials}
\noindent
Rigorous investigation of the complete general relativistic
multi-transonic BH accretion disc
structure is extremely complicated. 
At the same time it is
understood that, as relativistic effects play an important role in the
regions close to the accreting black hole (where most of the
gravitational potential energy is released), purely Newtonian gravitational
potential (in the form ${\Phi}_{N}=-\frac{1}{r}$)
cannot be a realistic choice to describe
transonic black hole accretion in general. To compromise between the ease of
handling of a
Newtonian description of gravity and the realistic situations
described by complicated general relativistic calculations, a series of
`modified' Newtonian potentials have been introduced
to describe the general relativistic effects that are
most important for accretion disk structure around Schwarzschild and Kerr
black holes (see ABN for further discussion).
 Introduction of such potentials allows one to investigate the
complicated physical processes taking place in disc accretion in a
semi-Newtonian framework by avoiding pure general relativistic calculations
so that
most of the features of spacetime around a compact object are retained and
some crucial properties of the analogous relativistic
solutions of disc structure could be reproduced with high accuracy.
Hence, those potentials might be designated as `pseudo-Kerr' or `pseudo-
Schwarzschild' potentials, depending on whether they are used to mimic the
space time around a rapidly rotating or non rotating/ slowly rotating
(Kerr parameter $a\sim0$) black
holes respectively.
Below we describe four such pseudo Schwarzschild potentials on
which we will concentrate in this paper.
It is important to note that 
as long as one is not
interested in astrophysical processes extremely close
(within $1-2~r_g$) to a black hole horizon, one may safely
use the following BH potentials to study
accretion on to a Schwarzschild
black hole with the advantage that use of these
potentials would simplify calculations by allowing one
to use some basic features of flat geometry
(additivity of energy or de-coupling of various
energy components, i.e., thermal ($\frac{a^2}{\gamma-1}$),
Kinetic ($\frac{u^2}{2}$) or
gravitational ($\Phi$) etc., see subsequent discussions)
which is not possible for
calculations in a purely Schawarzschild metric (see \S 4).
Also, one
can study more complex many body problems such as
accretion from an ensemble of companions or overall
efficiency  of accretion onto an ensemble of black holes
in a galaxy  or  for studying numerical hydrodynamic accretion flows
around a black hole etc. as simply as can be done in a
Newtonian framework, but with far better
accuracy. So we believe that a comparative study of multi-transonic 
accretion flow as well as shock formation using all these 
potentials might be quite useful in understanding some 
important features of various shock related astrophysical phenomena,
at least until one can have a complete and self-consistent 
theory of complete general relativistic shock formation exclusively
for a Schwarzschild BH.  However, one should be careful in using these
potentials because none of these potentials discussed here
are `exact' in a sense that they are not directly
derivable from the Einstein equations.
These potentials
could only be used to obtain more
accurate correction terms over and above the pure
Newtonian results and any `radically' new results
obtained using these potentials should be cross-checked
very carefully with the exact general relativistic theory.\\ 
\noindent
Paczy\'nski and Wiita (1980) proposed a pseudo-schwarzschild
potential of the form
$$
\Phi_{1}=-\frac{1}{2(r-1)}
\eqno{(1a)}
$$
which accurately reproduces the positions of $r_s$ and $r_b$ 
and gives the value 
of efficiency to be $-0.0625$ which is in closest agreement 
with the value obtained in full general relativistic calculations. 
Also the Keplarian distribution of angular 
momentum obtained using this potential is exactly same as 
that obtained in pure 
Schwarzschild geometry. 
It is worth mentioning here that this potential 
was first introduced to study a thick accretion disc with super Eddington 
Luminosity. Also,
it is interesting to note that although it had been thought of
in 
terms of disc accretion, $\Phi_1$ 
is spherically symmetric with a scale shift of 
$r_g$. \\
\noindent
To analyze the normal modes of accoustic oscillations within a 
thin accretion 
disc around a compact object (slowly rotating black hole or weakly 
magnetized neutron star), Nowak and Wagoner (1991) approximated some of the 
dominant relativistic effects of the accreting 
black hole (slowly rotating or 
nonrotating) via a modified Newtonian potential of the form
$$
\Phi_{2}=-\frac{1}{2r}\left[1-\frac{3}{2r}+12{\left(\frac{1}{2r}\right)}^2\right
]
\eqno{(1b)}
$$
$\Phi_2$ has correct form of $r_s$ as in the Schwarzschild case
but is unable to 
reproduce the value of $r_b$.
 This potential has the correct general relativistic value of the
angular velocity $\Omega_s$ 
at $r_s$. Also it reproduces the
radial epicyclic frequency $\kappa$ (for $r>r_s$) close to its value obtained
from general relativistic calculations, 
and among all BH potentials, $\Phi_2$ provides the best approximation for
$\Omega_s$ and $\kappa$.
However, this potential gives the
value of efficiency as $-0.064$ which is larger than that produced by 
$\Phi_1$, hence the disc spectrum computed using $\Phi_2$ would be more 
luminous compared to a disc structure studied using $\Phi_1$.\\
\noindent
Considering the fact that the free-fall acceleration plays a very crucial 
role in Newtonian gravity, ABN proposed two different 
BH  potentials to study disc accretion around a non-rotating black hole.
The first potential proposed by them produces exactly the
same value of the free-fall
acceleration of a test particle at a given value of $r$ as is obtained
for a test particle at rest with respect to the Schwarzschild reference
frame, and is given by
$$
\Phi_{3}=-1+{\left(1-\frac{1}{r}\right)}^{\frac{1}{2}}
\eqno{(1c)}
$$
The second one gives the value of the free fall acceleration that is equal 
to the value of the covariant component of the three dimensional free-fall 
acceleration vector of a test particle that is at rest in the Schwarzschild 
reference frame and is given by
$$
\Phi_{4}=\frac{1}{2}ln{\left(1-\frac{1}{r}\right)}
\eqno{(1d)}
$$
Efficiencies produced by $\Phi_3$ and $\Phi_4$ are $-0.081$ and $-0.078$ 
respectively.The magnitude of efficiency produced by $\Phi_3$
being 
maximum,calculation of disc structure using $\Phi_3$
will give  the maximum 
amount of energy dissipation and the corresponding spectrum would be the 
most luminous one. 
Hereafter we will refer to 
all these four potentials by $\Phi_i$ in 
general, where $\left\{i=1,2,3,4\right\}$ would correspond to $\Phi_1$
(eqn. 1(a)), $\Phi_2$ (eqn. 1(b)), $\Phi_3$ (eqn. 1(c)) and $\Phi_4$ (eqn. 1(d))
respectively.
One should notice that while all other $\Phi_i$s have
singularity at $r=r_g$, only $\Phi_2$ has a singularity at $r=0$.
It can be shown that for $r>2r_g$, while $\Phi_2$ is flatter
compared to purely Newtonian potential $\Phi_N$, all
other $\Phi_i$s are steeper to $\Phi_N$.\\
\noindent
At any radial distance $r$ measured from the accretor,
one can define the effective potential $\Phi_i^{eff}(r)$
to be the summation of the gravitational
potential and the centrifugal potential for matter
accreting under the influence of $i$th pseudo
potential. $\Phi_i^{eff}(r)$ can be expressed as:
$$
\Phi_i^{eff}(r)=\Phi_i(r)+\frac{\lambda^2(r)}{2r^2}
\eqno{(2a)}
$$
where $\lambda(r)$ is the non-constant distance dependent
specific angular momentum of accreting material. One
then easily shows that $\lambda(r)$ may have an upper limit:
$$
\lambda^{up}_i(r)=r^{\frac{3}{2}}\sqrt{\Phi^{'}_i(r)}
\eqno{(2b)}
$$
where $\Phi^{'}_i(r)$ represents the derivative of $\Phi_i(r)$ with
respect to $r$.
For weakly viscous or inviscid flow, angular
momentum can be taken as a constant parameter ($\lambda$) and eqn. (2a)
can be approximated as:
$$
\Phi_i^{eff}(r)=\Phi_i(r)+\frac{\lambda^2}{2r^2}
\eqno{(2c)}
$$
For general relativistic treatment of accretion, the
effective potential can {\it not} be decoupled in to its
gravitational and centrifugal components. 
For a Schwarzschild metric of the following form:
$$
ds^2=g_{{\mu}{\nu}}dx^{\mu}dx^{\nu}
$$
$$
=-\left(1-\frac{1}{r}\right)dt^2
+\left(1-\frac{1}{r}\right)^{-1}dr^2
+r^2\left(d{\theta}^2+sin^2{\theta}d{\phi}^2\right)
$$
the world line of the accreting fluid is timelike, and
the four velocity of the fluid satisfies the
normalization condition:
$$
u_{\mu}u^{\mu}=-1
$$
where $u^{\mu}$($u_{\mu}$) is the contra(co)-variant four velocity of the 
fluid. The angular velocity $\Omega$ of the fluid can be computed as
$$
\Omega=\frac{u^{\phi}}{u^t}=-\frac{{\lambda}g_{tt}}{g_{{\phi}{\phi}}}
=\frac{{\lambda}\left(r-1\right)}{2r^3}
$$
where $\lambda=-\frac{u_{\phi}}{u_t}$ is the specific angular momentum
which is conserved for fluid dynamics as well as for
particle dynamics for invicid flow. The  general
relativistic effective potential $\Phi^{eff}_{GR}(r)$ ({\it excluding}
the rest
mass) experienced by the fluid accreting on to a Schwarzschild BH 
can be expressed as:
$$
\Phi^{eff}_{GR}(r)=r\sqrt{\frac{r-1}{r^3-{\lambda}^2\left(1+r\right)}}-1
\eqno{(2d)}
$$
One can understand that the effective potentials in
general relativity cannot be obtained by linearly combining its
gravitational and rotational contributions because
various energies in general relativity are combined together to produce
non-linearly coupled new terms.\\
\noindent
In Fig 1, we plot $\Phi_i^{eff}(r)$ (obtained from eq. (2c)) and 
$\Phi^{eff}_{GR}(r)$ as a function of $r$ in logarithmic scale. The value of
$\lambda$ is taken to be 2
in units of $2GM/c$. $\Phi^{eff}$ curves for different $\Phi_i$
are marked exclusively in the
figure and the curve marked by ${\bf G^R}$ represents the
variation of $\Phi^{eff}_{GR}(r)$ with $r$. 
One can observe that $\Phi^{eff}_1(r)$ is in
excellent agreement with $\Phi^{eff}_{GR}(r)$, 
only for a very small
value of $r$ ($r{\rightarrow}r_g$),
$\Phi^{eff}_1$
starts deviating from $\Phi^{eff}_{GR}(r)$ and this deviation keeps
increasing as matter approaches closer and closer to
the event horizon. All other $\Phi^{eff}_i(r)$s
approaches to $\Phi^{eff}_{GR}(r)$ at a
radial distance (measured from the BH) considerably
larger compared to the case for $\Phi^{eff}_1(r)$. If one defines 
${\Delta}_i^{eff}(r)$
to be the measure of the deviation of $\Phi^{eff}_i(r)$ with
$\Phi^{eff}_{GR}(r)$
at any
point $r$, 
$$
{\Delta}_i^{eff}(r)=\Phi^{eff}_i(r)-\Phi^{eff}_{GR}(r)
$$
One observes that ${\Delta}_i^{eff}(r)$ is always negative for 
$\Phi^{eff}_1(r)$,
but for other $\Phi^{eff}_i(r)$, it normally remains positive for
low values of $\lambda$ but may become negative for a very
high value of $\lambda$. If 
${{\vert}}{\Delta}_i^{eff}(r){{\vert}}$
be the modules or the absolute
value of ${\Delta}_i^{eff}(r)$, one can also see that, although only for a
very small range of radial distance very close to the event horizon, 
${\Delta}_3^{eff}(r)$ is maximum,
for the whole range of distance scale while $\Phi_1$ is the
best approximation of general relativistic space time,
$\Phi_2$ is the worst approximation and $\Phi_4$ and $\Phi_3$ are the
second and the third best approximation as long as the
total effective potential experienced by the accreting
fluid is concerned. It can be shown that 
${{\vert}}{\Delta}_i^{eff}(r){{\vert}}$ nonlinearly
anti-correlates with $\lambda$. The reason behind this is
understandable. As $\lambda$ decreases, rotational mass as
well as its coupling term with gravitational mass
decreases for general relativistic accretion
material while for accretion in any $\Phi_i$, centrifugal
force becomes weak and gravity dominates; hence
deviation from general relativistic case will be more
prominent because general relativity is basically a
manifestation of strong gravity close to the compact
objects.\\
\noindent
From the figure it is clear that for $\Phi^{eff}_{GR}(r)$
as well as for
all $\Phi^{eff}_i(r)$s, a peak appears close to the horizon. The
height of these peaks may roughly be considered as the
measure of the strength of the centrifugal barrier
encountered by the accreting material for respective
cases. The deleberate use of the word `roughly' instead of
`exactly' is due to the fact that here we are dealing
with fluid accretion, and unlike particle dynamics, the
distance at which the strength of the centrifugal
barrier is maximum, is located further away from the
peak of the effective potential because here the total
pressure contains the contribution due to fluid or
`ram' pressure also. Naturally the peak height for  $\Phi^{eff}_{GR}(r)$
as well as for $\Phi^{eff}_i(r)$s increases with increase of $\lambda$ and
the location of this barrier moves away from the BH
with higher values of angular momentum. If the
specific angular momentum of accreting material lies
between the marginally bound and marginally stable
value, an accretion disc is formed. For invicid or
weakly viscous flow, the higher will be the value of $\lambda$,
the higher will be the strength of the centrifugal
barrier and the more will be the amount of radial
velocity or the thermal energy that the accreting material 
must have to begin with so that it can be made to accrete
on to the BH. In this connection it is important to
observe from the figure that accretion under $\Phi_1(r)$ will
encounter a centrifugal barrier farthest away from the
BH compared to other $\Phi_i$s.  For accretion under all $\Phi_i$s
except $\Phi_1$,the strength of centrifugal barrier at a
particular distance will be more compared to its value
for full general relativistic accretion.
\section{Multi-transonic flow in various BH potentials and shock 
formation}
\noindent
Following standerd literature, we consider a thin, 
rotating, axisymmetric, invicid steady flow in hydrostatic 
equilibrium in transverse direction. The assumption of hydrostatic
equilibrium is justified for a thin flow because for such flows, the infall
time scale is expected to exceed the local sound crossing time 
scale in the direction transverse to the flow. The flow is also assumed to 
posses considerably large radial velocity which makes the flow `advective'.
The complete solutions of such a system require the dimensionless
equations for conserved specific energy ${\cal E}$ and angular 
momentum $\lambda$ of the accreting material, the mass conservation 
equations supplied by the transonic conditions at the sonic points and the 
Rankine Hugoniot conditions at the shock. The local half-thickness,
$h_i(r)$ of the disc for any $\Phi_i(r)$ can be obtained by balancing the
gravitational force by pressure gradient and can be expressed as:
$$
h_i(r)=a\sqrt{{r}/\left({\gamma}{\Phi_i^{\prime}}\right)}
$$
For a non-viscous flow obeying the polytropic equation of state
$p=K{\rho}^{\gamma}$ ($K$ is a measure of
the specific entropy of the flow),  integration of radial momentum
equation:
$$
u\frac{{d{u}}}{{d{r}}}+\frac{1}{\rho}
\frac{{d}P}{{d}r}+\frac{d}{dr}\left(\Phi^{eff}_{i}(r)\right)=0
$$
leads to the following energy conservation equation in steady state:
$$
{\cal E}=\frac{1}{2}u_e^2+\frac{a_e^2}{\gamma - 1}
+\frac{{\lambda}^2}{2r^2}+\Phi_i=0;
\eqno{(3a)}
$$
and the continuity equation:
$$
\frac{{d}}{{d}r}
\left[u{\rho}rh_i(r)\right]=0
$$
can be integrated to obtain the barion number conservation equation:
$$
{\dot M}_{in}=\sqrt{\frac{1}{\gamma}}u_ea_e{\rho}_er^{\frac{3}{2}}
\left({\Phi_i^{\prime}}\right)^{-\frac{1}{2}}.
\eqno{(3b)}
$$
Following C89, one can define the entropy accretion rate
${\dot {\cal M}}={\dot M}_{in}K^{\left(\frac{1}{\gamma-1}\right)}
{\gamma}^{\left(\frac{1}{\gamma-1}\right)}$
which undergoes a  discontinuous transition
at the shock location $r_{sh}$ where local turbulence generates entropy to
increase ${\dot {\cal M}}$ for post-shock flows. For our purpose,
explicit expression for ${\dot {\cal M}}$ can be obtained as:
$$
{\dot {\cal M}}=\sqrt{\frac{1}{\gamma}}u_e
a_e^{\left({\frac{\gamma+1}{\gamma-1}}\right)}
r^{\frac{3}{2}}\left({\Phi_i^{\prime}}\right)^{-\frac{1}{2}}.
\eqno{(3c)}
$$
In Eqs. (3a-3c), the subscript $e$ indicates the values measured on the
equatorial plane of the disc;
however,
we will drop $e$ hereafter if no confusion arises in doing so.
One can simultaneously solve Eqs. (3a - 3c) 
for any particular $\Phi_i(r)$ and for a
particular set of values of $\left\{{\cal E}, \lambda, \gamma\right\}$.
Hereafter we will use the notation $\left[{\cal P}_i\right]$ for a set of
values of $\left\{{\cal E}, \lambda, \gamma\right\}$ for any particular
$\Phi_i$.\\
\noindent
For a particular value of  $\left[{\cal P}_i\right]$, it is now quite 
straight-forward to derive the space gradient of dynamical flow velocity
$\left(\frac{du}{dr}\right)_i$ for flow in any particular 
$i$th BH potential $\Phi_i(r)$:
$$
\left(\frac{du}{dr}\right)_i=
\frac{
\left(\frac{\lambda^2}{r^3}+\Phi^{'}_i(r)\right)-
\frac{a^2}{\gamma+1}\left(\frac{3}{r}+
\frac{\Phi^{''}_i(r)}{\Phi^{'}_i(r)}\right)
}
{u-\frac{2a^2}{u\left(\gamma+1\right)}
}
\eqno{(4a)}
$$
where
${\Phi_i}^{{\prime}{\prime}}$ represents the derivative
of ${\Phi_i}^{{\prime}}$.
Since the flow is assumed to be smooth everywhere, if
the denominator of eqn. 4(a)  vanishes at any radial distance
$r$, the numerator must also vanish there to maintain the
continuity of the flow. One therefore arrives at the so
called `sonic point (alternately, the `critical point')
conditions'  by simultaneously making
the numerator and denominator of eqn. 4(a) equal zero.
The sonic point conditions can be expressed as:
$$
a^i_s=\sqrt{\frac{1+\gamma}{2}}u^i_s=
\left[
\frac{\Phi^{'}_i(r)+{\gamma}\Phi^{'}_i(r)}{r^2}
\left(
\frac{\lambda^2+r^3\Phi^{'}_i(r)}{3\Phi^{'}_i(r)+r\Phi^{''}_i(r)}
\right)
\right]_s
\eqno{(4b)}
$$
where the subscript $s$ indicates that the quantities are to be measured
at the sonic point(s). For a fixed $\left[{\cal P}_i\right]$,
one can solve the following polynomial of $r$
to obtain the sonic point(s) of the flow:
$$
{\cal E}-{\left(\frac{\lambda^2}{2r^2}+\Phi_i
\right)}_{s}-\frac{2\gamma}{\gamma^2-1}
\left[
\frac{\Phi^{'}_i(r)+{\gamma}\Phi^{'}_i(r)}{r^2}
\left(
\frac{\lambda^2+r^3\Phi^{'}_i(r)}{3\Phi^{'}_i(r)+r\Phi^{''}_i(r)}
\right)
\right]_s
=0.
\eqno{(4c)}
$$
Similarly, the value of $\left(\frac{du}{dr}\right)_i$ at its 
corresponding sonic point(s) can be obtained by solving the 
following equation:
$$
\frac{4{\gamma}}{\gamma+1}\left(\frac{du}{dr}\right)^2_{s,i}
-2u_s\left(\frac{\gamma-1}{\gamma+1}\right)
\left(\frac{3}{r}+\frac{\Phi^{''}_i(r)}{\Phi^{'}_i(r)}\right)_s
\left(\frac{du}{dr}\right)_{s,i}
$$
$$
+a^2_s\left[\frac{\Phi^{'''}_i(r)}{\Phi^{'}_i(r)}
-\frac{2\gamma}{\left(1+{\gamma}\right)^2}
\left(\frac{\Phi^{''}_i(r)}{\Phi^{'}_i(r)}\right)^2
+\frac{6\left(\gamma-1\right)}{\gamma{\left(\gamma+1\right)^2}}
\left(\frac{\Phi^{''}_i(r)}{\Phi^{'}_i(r)}\right)
-\frac{6\left(2\gamma-1\right)}{\gamma^2{\left(\gamma+1\right)^2}}
\right]_s
$$
$$
+
\Phi^{''}_i{\Bigg{\vert}}_s-
\frac{3\lambda^2}{r^4_s}=0
\eqno{(4d)}
$$
Where the subscript $(s,i)$ indicates that the corresponding 
quantities for any $i$th potential is being measured at its 
corresponding sonic point(s) and $\Phi^{'''}_i(r)=\frac{d^3\Phi_i(r)}{dr^3}$.\\
\noindent
For {\it all} $\Phi_i$'s,
we find a significant region of
parameter space spanned by $\left[{\cal P}_i\right]$ which allows
the multiplicity of
sonic points for accretion as well as for wind
where two real physical inner and outer (with respect to
the BH location) $X$ type sonic points $r_{in}$ and $r_{out}$ encompass
one $O$ type unphysical middle sonic point $r_{mid}$ in between.
For a particular
$\Phi_i$, if
${\cal A}_i\left[{\cal P}_i\right]$ denotes the universal set
representing the entire parameter space covering all
values of $\left[{\cal P}_i\right]$, and if
${\cal B}_i\left[{\cal P}_i\right]$ represents one particular  subset
of
${\cal A}_i\left[{\cal P}_i\right]$
which contains  only the
particular values of $\left[{\cal P}_i\right]$ for which the above mentioned
three sonic points are obtained, then ${\cal B}_i\left[{\cal P}_i\right]$
can further be decomposed into two subsets ${\cal C}_i\left[{\cal P}_i\right]$
and ${\cal D}_i\left[{\cal P}_i\right]$ such that:
$$
{\cal C}_i\left[{\cal P}_i\right]~\subseteq~
{\cal B}_i\left[{\cal P}_i\right]~
{\rm \underline{only~for}}~
{\dot {\cal M}}\left(r_{in}\right)>
{\dot {\cal M}}\left(r_{out}\right),
$$
and
$$
{\cal D}_i\left[{\cal P}_i\right]~\subseteq~
{\cal B}_i\left[{\cal P}_i\right]~
{\rm \underline{only~for}}~
{\dot {\cal M}}\left(r_{in}\right)<
{\dot {\cal M}}\left(r_{out}\right),
$$
then for $\left[{\cal P}_i\right] \in {\cal C}_i\left[{\cal P}_i\right]$,
we get multi-transonic {\it accretion} and for
$\left[{\cal P}_i\right] \in {\cal D}_i\left[{\cal P}_i\right]$
one obtains  multi-transonic {\it wind}.
In Fig. 2. we plot
$$
\left({\cal E}_i,\lambda_i\right) \in \left[{\cal P}_i\right] 
\in {\cal C}_i\left[{\cal P}_i\right]~\subseteq~
{\cal B}_i\left[{\cal P}_i\right]~
{\rm \underline{and}}~
\left({\cal E}_i,\lambda_i\right) \in \left[{\cal P}_i\right] 
\in {\cal D}_i\left[{\cal P}_i\right]~\subseteq~
{\cal B}_i\left[{\cal P}_i\right]
$$
for all $\Phi_i(r)$s (marked in the figure) when $\gamma=4/3$. While the 
specific energy ${\cal E}$ is plotted along the $Y$ axis, the 
specific angular momentum $\lambda$ is plotted along $X$ axis. For 
$\Phi_1(r)$, the shaded region {\bf PQR} represents the parameter space
spanned by ${\cal E}$ and $\lambda$ for which
three sonic points will form in {\it accretion} ({\bf PQR}${\equiv}
{\cal C}_1\left[{\cal P}_1\right]$)
while the wedge shaped un-shaded region {\bf PSR} represents the 
parameter space for which three sonic points are formed in {\it wind}
({\bf PSR}${\equiv}{\cal D}_1\left[{\cal P}_1\right]$). Similar 
kind of parameter space division is shown for other $\Phi_i(r)$s as well.
A careful analysis of Fig. 2 reveals the fact that, at least 
for ultra-relativistic flow, {\it no} region of parameter space
common to all $\Phi_i(r)$ 
is found for which $\left[{\cal P}_i\right]\in{\cal C}_i\left[{\cal P}_i\right]$
or $\left[{\cal P}_i\right]\in{\cal D}_i\left[{\cal P}_i\right]$. 
However, significant region of parameter space is obtained for which 
$\left[{\cal P}_i\right]\in{\cal C}_i\left[{\cal P}_i\right]$
or $\left[{\cal P}_i\right]\in{\cal D}_i\left[{\cal P}_i\right]$ for $\Phi_2(r)$
and $\Phi_3(r)$ and a very small region of such common zone in the 
parameter space is 
obtained (only for extremely low values of the energy and angular 
momentum of the acceting matter) for $\Phi_2(r)$, $\Phi_3(r)$ and $\Phi_4(r)$.
As the flow approaches to its purely non-relativistic limit, i.e.,
as we make $\gamma~{\longrightarrow}~5/3$, tendency for such mutual overlap
of parameter space for $\Phi_2(r)$, $\Phi_3(r)$ and $\Phi_4(r)$
increases. Nevertheless, $\Phi_1$ still remains 
`untouchable' by $\Phi_2(r)$ and $\Phi_3(r)$; only a particular region of
parameter space (fairly low energy accretion with intermediate value
of angular momentum) is commonly shared by $\Phi_4(r)$ and $\Phi_1(r)$.\\
\noindent
One also observe that if ${\cal E}_i^{max}$ and  ${\lambda}_i^{max}$
are the maximum available energy and angular momentum of the flow
for any $\Phi_i(r)$ for which 
$\left[{\cal P}_i\right] \in {\cal C}_i\left[{\cal P}_i\right]$
or $\left[{\cal P}_i\right] \in {\cal D}_i\left[{\cal P}_i\right]$,
one can write:
$$
{\cal E}_3^{max}~>~{\cal E}_4^{max}~>~{\cal E}_1^{max}~>~
{\cal E}_2^{max}
$$
and
$$
{\lambda}_1^{max}~>~{\lambda}_4^{max}~>~{\lambda}_2^{max}~>~
{\lambda}_3^{max}
$$
The above trend remains unaltered as $\gamma~{\longrightarrow}~5/3$ and we 
observe that both  ${\cal E}_i^{max}$ and  ${\lambda}_i^{max}$
non-linearly anti-correlates with $\gamma$.\\
\noindent
If shock forms in accretion (in this work we will not study the 
shock formation in wind), then $\left[{\cal P}_i\right]$s responsible 
for shock formation must be somewhere from the region for which 
$\left[{\cal P}_i\right] \in {\cal C}_i\left[{\cal P}_i\right]$, though 
not all $\left[{\cal P}_i\right] \in {\cal C}_i\left[{\cal P}_i\right]$
will allow shock transition.
Using Eqs. (3a - 3c),
we combine the three standard Rankine-Hugoniot
conditions (Landau \& Lifshitz 1959)
for vertically integrated pressure and density
(see Matsumoto et al. 1984)
to
derive the following relation which is valid {\it only} at the shock
location: \\ \\
$$
\left(1-\gamma\right)\left(\frac{{\rho_{-}}{{\dot {\cal M}}_{-}}}{\dot M}
\right)^{log_{\Gamma}^{1-\Theta}}
{\cal E}_{{\left(ki+th\right)}}
-{\Theta}{\left(1+\Theta-R_{comp}\right)}^{-1}
+\left(1+\Theta\right)^{-1}
=0,
\eqno{(5)}
$$
where ${\cal E}_{{\left(ki+th\right)}}$ is the total specific thermal plus
mechanical energy of the accreting fluid:
${\cal E}_{{\left(ki+th\right)}}=\left[{\cal E}-
\left(\frac{\lambda^2}{2r^2}+\Phi_i\right)\right]$, $R_{comp}$ and $\beta$ are
the density compression and entropy enhancement ratio respectively, defined
as
$R_{comp}=\left({\rho_{+}}/{\rho_{-}}\right)$ and
$\beta=\left({\dot {\cal M}}_{+}/{\dot {\cal M}}_{-}\right)$
respectively; $\Theta=1-\Gamma^{\left(1-{\gamma}\right)}$ and $\Gamma={\beta}
{R_{comp}}$, ``$+$'' and ``$\_$'' refer to the post- and
pre-shock quantities.
The shock
strength ${\cal S}_i$ (ratio of the pre- to post-shock Mach number of the
flow) can be calculated as:
$$
{\cal S}_i=R_{comp}\left(1+\Theta\right).
\eqno{(6)}
$$
Eqs. (5) and (6) cannot be solved
analytically because they are non-linearly coupled. However, we have been
able to simultaneously solve Eqs.\ (3 - 6) using iterative numerical
techniques. We have developed an efficient numerical
code which takes
$\left[{\cal P}_i\right]$ and $\Phi_i$ as its input and can calculate $r_{sh}$
along with
any sonic or shock quantity as a function of
$\left[{\cal P}_i\right]$.
It is to be noted that like the references cited in \S 1, we also obtain
multiplicity in the shock location. Following C89, we perform
the local stability analysis and find that only one $r_{sh}$ which forms in
between $r_{out}$ and $r_{mid}$ is
stable for {\it all} $\Phi_i$.\\
\noindent
If $\left[{\cal P}_i\right]
\in {\cal F}_i\left[{\cal P}_i\right]~\subseteq~
{\cal C}_i\left[{\cal P}_i\right]$
represents the region of parameter space for which 
multi-transonic supersonic
flows is expected to
encounter a RHS at $r_{sh}$, where they
become hotter, shock compressed and subsonic
and will again become supersonic only after passing through $r_{in}$ before
ultimately crossing the event horizon, then one can also define
$\left[{\cal P}_i\right]
\in {\cal G}_i\left[{\cal P}_i\right]$ which is complement 
of ${\cal F}_i\left[{\cal P}_i\right]$ related to 
${\cal C}_i\left[{\cal P}_i\right]$ so that for:
$$
\left\{{\cal G}_i\left[{\cal P}_i\right]\Bigg{\vert}
\left[{\cal P}_i\right]\in{\cal C}_i\left[{\cal P}_i\right]
~{\rm and}~
\left[{\cal P}_i\right]\notin{\cal F}_i\left[{\cal P}_i\right]
\right\},
$$
the shock location becomes imaginary in
${\cal G}_i\left[{\cal P}_i\right]$,
hence no stable RHS forms in that region;
rather the shock keeps oscillating back and forth. We anticipate that
${\cal G}_i\left[{\cal P}_i\right]$ is also an important zone which might be
responsible for the Quasi-Periodic Oscillation (QPO) of the BH candidates
(see \S 5).\\
\noindent
Figure 3 demonstrates few typical flow topologies of the integral curves
of motion for ultra-relativistic ($\gamma=4/3$)
shocked flows in various $\Phi_i$'s
(indicated in the figure).
While the distance from the event horizon of the central BH 
(scaled in the units of $r_g$ and plotted in logarithmic scale) is 
plotted along the X axis, the local  Mach number of the flow is plotted 
along the Y axis.
One can easily obtain such a set of figures for any $\gamma$
(and $\left[{\cal P}_i\right]$) which allow shock formation.
For all figures, {\bf ABCD} represents
the transonic accretion passing through the outer sonic point $r_{out}$
(marked as {\bf B}) if a shock would not form. However, as ${\dot {\cal M}}$
of the flow is higher at the inner sonic point $r_{in}$ compared to
${\dot {\cal M}}$ at $r_{out}$, the flow must encounter a 
shock at {\bf C} (the vertical
line {\bf CE} marked by an arrowhead represents the shock transition), becomes
subsonic and jumps on the branch {\bf EF}, which ultimately hits the event horizon
supersonically after it passes through the inner sonic point $r_{in}$,
which is
marked on {\bf EF} by the small circle with a dot at the center.
An ``$\ast$" in the figure indicates the location of the middle sonic point
$r_{mid}$. The corresponding values of $r_{in}$, $r_{mid}$, $r_{out}$,
the shock location $r_{sh}$, and shock strength $S_i$ are indicated at the top
of each figure, while the corresponding values of the total specific energy
${\cal E}$ and angular momentum $\lambda$ for which the solutions are obtained,
are indicated inside each figure. {\bf GBH} represents the `self-wind' of the
flow, which, in the course of its motion {\it away} from the BH to infinity,
becomes supersonic after passing through $r_{out}$ at {\bf B}. Collectively,
{\bf ABCEF} represents the real physical shocked accretion which connects
infinity with the event horizon.
The overall scheme for obtaining the above mentioned integral curves is 
as follows:\\
\noindent
First we compute $r_{in}$, $r_{mid}$ and $r_{out}$ by solving eq. (4c). Then
we obtain the dynamical velocity gradient of the flow at sonic points
by solving eq. (4d). For a chosen ${\dot M}_{in}$ (scaled in the units of
the Eddington rate ${\dot M}_{Edd}$), we then compute the local dynamical 
flow velocity $u(r)$, the local polytropic sound speed $a(r)$, the 
local radial Mach number $M(r)$, the local fluid density ${\rho}(r)$
and any other related dynamical or thermodynamic quantities by solving the
eq. (4a-4d) from the outer sonic point using fourth order Runge-Kutta method.
We start integrating from $r_{out}$ in two different directions. Along
{\bf BH}, we only solve for $u(r)$, $a(r)$ and $M(r)$ because shock does not
form in subsonic flows. However, integration along {\bf BCD} involves a
different procedure. Along {\bf BCD}, not only we compute 
$u(r)$, $a(r)$ and $M(r)$, but also, at {\it every} integration step 
(with as small a step size as possible), we keep checking 
whether eq. (5) is being satisfied at that point. To do so, at each 
and every point, we start with a suitable initial guess value of $R_{comp}$
and ${\cal S}_i$ and performs millions of iteration to check whether 
for any set of $\left[R_{comp},{\cal S}_i\right]$, eq. (5) is satisfied at that
point and whether for such $\left[R_{comp},{\cal S}_i\right]$, the value
of $\beta$ obtained from eq. (5) becomes exactly equal to 
$\frac{{\dot {\cal M}}(r_{in})}{{\dot {\cal M}}(r_{out})}$; in other words,
whether the entropy generated at that point (if any) becomes exactly equal 
to the difference between the entropies at the inner and the outer sonic 
points respectively. If such conditions are satisfied at some particular point
(point {\bf C} in the figure), we argue that the shock forms at that point
and we can calculate any pre- and the post-shock dynamical and thermodynamic
quantities at the shock location $r_{sh}$ (i.e., at {\bf C}). Once a 
shock is formed, the flow jumps from its supersonic branch {\bf BCD} to 
its subsonic branch {\bf EG}. We again start calculating 
$u(r)$, $a(r)$ and $M(r)$ and any other related flow quantities by solving
eq. (4a) using fourth order Runge-Kutta method (with the help of
eq. (3a-3c) and eq. (4d)), but this time from the {\it inner} sonic point
$r_{in}$ of the flow.\\
\noindent
In Figure
4, we present the ${\cal F}_i\left[{\cal P}_i\right]$s for
all four $\Phi_i$'s ($\Phi_1$ $\rightarrow$(a), 
$\Phi_2~\rightarrow$ (b), $\Phi_3~\rightarrow$ (c), and
$\Phi_4~\rightarrow$ (d)).
The specific energy ${\cal E}$, specific angular momentum $\lambda$ 
and the polytropic index $\gamma$ of the flow are plotted along the 
Z, Y and X axis respectively.
Each surface for a particular $\Phi_i(r)$ is drawn for a particular
value of $\gamma$. While the first
surfaces (which have the maximum surface areas)
on the $\left({\cal E}~-~\gamma\right)$ plane represent ultra-relativistic
accretion ($\gamma=4/3$), successive surfaces are also shown for higher values
of $\gamma$, taking a regular interval of ${\Delta}\gamma=0.025$. It is 
observed that as the flow approaches to its purely non-relativistic limit, 
the area of the $\left({\cal E}~-~\gamma\right)$ surfaces responsible
for shock formation starts shrinking.
We find that the shock location  correlates with
$\lambda$. 
This is obvious because the higher will be the flow 
angular momentum, the greater will be the rotational energy content 
of the flow and the higher will be the strength of the centrifugal 
barrier (which is responsible to break the incoming flow by forming a shock)
as well as the further will be the location of such barrier from the 
event horizon.
However, $r_{sh}$ anti-correlates with ${\cal E}$ and $\gamma$.
which means that for same ${\cal E}$ and $\lambda$, shock in purely 
non-relativistic flow will form closer to the event horizon compared to
the ultra-relativistic flow. We also observe that the shock strength 
${\cal S}_i$ non-linearly anti-correlates with the shock location $r_{sh}$,
which indicates that 
the
closer the shock forms to the BH, the higher is the strength ${\cal S}_i$
and the entropy enhancement ratio $\beta$. The ultra-relativistic flows
are supposed to  
produce the strongest shocks. 
The reason behind this is also easy to understand; the closer the shock
forms to the event horizon, the higher will be the available gravitational 
potential energy to be released and the higher will be the radial 
advective velocity required to have a more vigorous shock jump.
Compared to $\Phi_2$ and
$\Phi_3$, $\Phi_1$ and $\Phi_4$ allow  wider spans of $\gamma$ as well
as $\lambda$ for shock formation. If ${\cal E}_{max}$, ${\lambda}_{max}$ and
$\gamma_{max}$ represents the maximum values of the corresponding parameters
for which shock formation is possible, we obtain
${\cal E}_{max}\left(\Phi_3\right)>
{\cal E}_{max}\left(\Phi_4\right)>
{\cal E}_{max}\left(\Phi_1\right)>
{\cal E}_{max}\left(\Phi_2\right)$,
${\lambda}_{max}\left(\Phi_1\right)>
{\lambda}_{max}\left(\Phi_4\right)>
{\lambda}_{max}\left(\Phi_3\right)>
{\lambda}_{max}\left(\Phi_2\right)$
and
${\gamma}_{max}\left(\Phi_4\right)>
{\gamma}_{max}\left(\Phi_1\right)>
{\gamma}_{max}\left(\Phi_3\right)>
{\gamma}_{max}\left(\Phi_2\right)$, respectively.
Also we observe that as more and more the flow approaches its purely 
non-relativistic limit,
shock may form for less and less angular momentum. For some
$\Phi_i(r)$s, even a very small amount of angular momentum ($\lambda<1$)
allows shock formation, which indicates that for purely non-relativistic
accretion, shock formation may take place even for quasi-spherical flow.
\section{General Relativistic multi-transonic accretion}
\noindent
Following the arguments provided by NT and
Chakrabarti 1996a, we derive the 
expressions for the conserved total specific energy ${\cal E}^{\prime}$
(which {\it includes} the rest mass energy) and the entropy accretion rate 
${\dot {\cal M}}$ as:
$$
{\cal E}^{\prime}=\frac{{\gamma}\left(\gamma-1\right)}
{\gamma-\left(1+a^2\right)}
\sqrt{\frac{r-1}{1-u^2}}
\left[r^3+{\lambda}^2\left(1-r\right)\right]^{-\frac{1}{2}}
\eqno{(7a)}
$$
and 
$$
{\dot {\cal  M}}=5.657ur^{1.25}\sqrt{\frac{r-1}{1-u^2}}
\left[\frac{a^2\left(\gamma-1\right)}{\gamma-\left(1+a^2\right)}\right]
^{\frac{\gamma+1}{2\left(\gamma-1\right)}}
\left[r^3+{\lambda}^2\left(1-\gamma\right)\right]^{0.25}
\eqno{(7b)}
$$
One can see from eq. (7a) that the total specific  energy in this case,
can {\it not} be decoupled into various linearly additive contributions 
of separate physical origin (i.e., kinetic, thermal, rotational or 
gravitational) as it could be done for flows in any pseudo-potential.\\
\noindent
Following the procedure outlined in previous section, one can derive
the dynamical flow velocity gradient for general relativistic accretion flow 
as:
$$
\left(\frac{du}{dr}\right)=
\left(
\frac
{
\frac{1}{2r}\left[\frac{2r^3-\lambda^2}{r^3+\lambda^2\left(1-\gamma\right)}
\right]
-\frac{2r-1}{2r\left(r-1\right)}
-\frac{2a^2}{\gamma+1}
\left[\frac{5-7r}{4r\left(r-1\right)}
+\frac{\lambda^2-3r^2}{4\left[r^3+\lambda^2\left(1-r\right)\right]}
\right]
}
{\left[\frac{2a^2}{u\left(u^2-1\right)\left(\gamma+1\right)}
+\frac{u}{1-u^2}\right]}
\right)
\eqno{(8a)}
$$
from which the sonic point conditions comes out to be:
$$
u_s=\sqrt{\frac{2}{\gamma+1}}a_s=
\sqrt{
\frac{\gamma+1}{2}
\left[
\frac{
\frac{1}{2r}\left[\frac{2r^3-\lambda^2}{r^3+\lambda^2\left(1-\gamma\right)}
\right]
-\frac{2r-1}{2r\left(r-1\right)}
}
{\frac{5-7r}{4r\left(r-1\right)}
+\frac{\lambda^2-3r^2}{4\left[r^3+\lambda^2\left(1-r\right)\right]}
}
\right]_s}
\eqno{(8b)}
$$
The sonic point(s) could be computed by solving the following 
equation:
$$
{{\cal  E}^{\prime}}^2\left[r_s^3+\lambda^2\left(1-r_s\right)\right]
-\frac{r_s-1}{1-{\bf {{\Psi}\left(r_s,\lambda\right)}}}
\left[\frac{\gamma\left(\gamma-1\right)}{\gamma-
{\bf {{\eta}\left(r_s,\lambda\right)}}}\right]^2=0
\eqno{(8c)}
$$
where
$$
{\bf {{\eta}\left(r_s,\lambda\right)}}
=\left[1+\frac{\gamma+1}{2}{\bf {{\Psi}\left(r_s,\lambda\right)}}\right]
$$
and
$$
{\bf {{\Psi}\left(r_s,\lambda\right)}}=
\left[
\frac{
\frac{1}{2r_s}\left[\frac{2r_s^3-\lambda^2}{r_s^3+\lambda^2\left(1-\gamma\right)}
\right]
-\frac{2r_s-1}{2r_s\left(r_s-1\right)}
}
{\frac{5-7r_s}{4r\left(r_s-1\right)}
+\frac{\lambda^2-3r_s^2}{4\left[r_s^3+\lambda^2\left(1-r_s\right)\right]}
}
\right]
$$
The dynamical flow velocity gradient at the sonic point(s) can be obtained
by solving the following equation:
$$
\frac{2\left(2\gamma-3a_s^2\right)}
{\left(\gamma+1\right)\left(u_s^2-1\right)^2}
\left(\frac{du}{dr}\right)_s^2+
4{\bf {\xi\left(r_s,\lambda\right)}}
\left[\frac{\gamma-\left(1+a_s^2\right)}{u_s^2-1}\right]
\left(\frac{du}{dr}\right)_s
$$
$$
+ \frac{2}{\gamma+1}a_s^2{\bf {\xi\left(r_s,\lambda\right)}}
{\Bigg [}
2{\bf {\xi\left(r_s,\lambda\right)}}\left[
\frac{\gamma-\left(1+a_s^2\right)}{\gamma+1}\right]-
\frac{2r_s-1}{r_s\left(r_s-1\right)}
-\frac{3r_s^2-\lambda^2}{r_s^3+\lambda^2\left(1-r_s\right)}
$$
$$
+\frac{40r_s^3-24r_s^2-\lambda^2\left(16r_s-13\right)}
{10r_s^4-8r_s^3-\lambda^2\left(8r_s^2-13r_s+5\right)}
{\Bigg  ]}=0
\eqno{(8d)}
$$
where
$$
{\bf {\xi\left(r_s,\lambda\right)}}=
\left[
\frac{5-7r}{4r\left(r-1\right)}
+\frac{\lambda^2-3r^2}{4\left[r^3+\lambda^2\left(1-r\right)\right]}
\right]
$$
We solve eq. (8c) and find that like flows in various $\Phi_i(r)$s,
here also a significant region of parameter space allows the multiplicity 
of sonic points for accretion as well as for wind where one $O$ type
unphysical middle sonic point is flanked in between 
two $X$ type real physical sonic
points $r_{in}$ and $r_{out}$ respectively. In Fig. 5 we show the 
regions of parameter space for which multi-transonic flow is
obtained for both accretion and wind. The dimensionless 
conserved total specific energy ${\cal E}$ ({\it excluding} the rest 
mass energy) is plotted along the Y axis whereas the specific angular 
momentum $\lambda$ is plotted along the X axis. 
In region bounded by {\bf PQR} and marked by ${\cal A}$, three sonic points are 
formed in accretion and in region bounded by {\bf PRS} and marked by 
${\cal W}$, three sonic points are formed in wind. While the figure is drawn
for ultra-relativistic flows, the corresponding regions of parameter space
can be obtained for any $\gamma$. If ${\cal E}_{max}$ be the maximum value 
of the energy and if $\lambda_{max}$ and $\lambda_{min}$ be the maximum and
minimum values of the angular momentum respectively, 
for which three
sonic points are formed in accretion for any particular $\gamma$, we observe that 
$\left[{\cal E}_{max},\lambda_{max},\lambda_{min}\right]$ non-linearly 
anti-correlates with $\gamma$. In other words, as the flow approaches its
purely non-relativistic limit, the area of the region involved in formation 
of multi-transonic accretion decreases to a lower value.\\
\noindent
In Fig. 6, we show the integral curves of motion for general relativistic 
accretion of ultra-relativistic polytropic fluid. For a particular  set
of $\left[{\cal E},\lambda,\gamma\right]$ shown in the figure, {\bf ABCD}
represents the accretion passing through the outer sonic point $r_{out}$
(marked in the figure by {\bf B}) location of which can be found by solving
eq. (8c). {\bf EBI} represents the self-wind. Flow along {\bf EFGH} passes
through the inner sonic point $r_{in}$ (marked in the figure by {\bf F}) and
encompasses a middle sonic point $r_{mid}$ location of which is shown in the figure 
using a ``$\ast$". Like Fig. 3, here also we obtain the complete solution
topology by integrating eq. (8a) (with the help of eq. (7a-7b) and eq. (8c))
using fourth order Runge-Kutta method.\\
\noindent
If $\Sigma$ and $\Pi$ be the shock compression and the entropy 
enhancement ratio (at the shock location) for this case 
($\Sigma=\frac{M_{-}}{M_{+}}$, $\Pi=\frac{{\dot {\cal M}}_{+}}{{\dot {\cal M}}_{-}}$),
one can show that the following equation will be satisfied when shock forms:
$$
\Pi{\Sigma}^
{\frac{1}{1-\gamma}}
\left(\frac{T_{-}}{T_{+}}\right)^{\frac{\gamma}{\gamma-1}}
\left(\frac{1-u_-^2}{1-u_{+}^2}\right)^
{\frac{1}{4}\left(\frac{3-\gamma}{\gamma-1}\right)}=1
\eqno{(9)}
$$
where $T(-/+)$ and $u(-/+)$ are the pre-/post-shock temperature and dynamical
velocities of the flow respectively. However, it is our
limitation in this paper that we have not been able to formulate or solve
any equation which can be used to calculate the shock location in general
relativistic accretion onto Schwarzschild BHs. Nevertheless, if shock forms
in such flow (which is, of course, expected), it is obvious that the set 
of $\left({\cal E},\lambda\right)$ responsible for shock formation
{\it must} belong to the region {\bf PQR} (${\equiv}$ $\pgrcgr$,
see \S 3) of Fig. 5 because shock will form {\it only} in 
multi-transonic accretion. The above argument is useful to compare 
accretion flows in various $\Phi_i(r)$s with general relativistic accretion
(at least as long as the question of of shock formation in multi-transonic flow
is concerned) in the following way:\\
\noindent
Suppose for ultra-relativistic flows,
we take the region of parameter space $\pifi$ for any
$\Phi_i(r)$ used in this paper (see Fig. 4), and then superpose that 
region with {\bf PQR} of Fig. 5 and study which $\Phi_i(r)$
provides the maximum overlap between $\pifi$ and $\pgrcgr$.
That particular BH potential is then considered to be the
most efficient pseudo-potential in approximating the general relativistic,
multi-transonic, shocked BH accretion. However, such an `efficiency test'
is not entirely unambiguous because as we are yet to figure out the
exact $\pgrfgr$, there may be some possibility that
for any $\Phi_i(r)$, though $\pifi$ will overlap with 
$\pgrcgr$, but instead of falling onto $\pgrfgr$, it will rather 
overlap with $\pgrggr$ because the exact boundary between 
$\pgrfgr$ and $\pgrggr$ could not be explored in our work.
Nevertheless, we believe that still our arguments for the 
`efficiency test' is of some use, at least until one can
find out the exact shock formation zone for general relativistic 
flow.\\
\noindent
In Fig. 7, we superpose the Fig. 5 on $\pifi$ for all 
different $\Phi(r)$s (marked in the figure) used in our work.
Unlike other $\pifi$s, $\pthfth$ are drawn using long-dashed
lines to show its overlap with $\ptwoftwo$. The figure is
drawn for ultra-relativistic flow but can also 
be drawn for other $\gamma$s as well. We observe that 
while $\ponefone$ has {\it excellent} overlap (except
at very high energy) with $\pgrcgr$, {\it no other}
$\pifi$s have any overlap with it. This leads to the conclusion that
at least for ultra-relativistic flow, not only $\Phi_1(r)$ is
a very good approximation, rather it is the {\it only} BH  potential to
approximate for the general relativistic multi-transonic shocked flow.
However, as the flow approaches its purely non-relativistic limit,
we observe that the area of the ovarlaping zone for $\Phi_1(r)$
decreases with higher $\gamma$ and $\ponefone$ is pushed back
to overlap rather with $\pgrdgr$; hence unlike ultra-relativistic accretion,
$\Phi_1(r)$ may not be
considered such a good approximation for purely non-relativistic 
flows. Also we find that a region of low energy - high angular momentum
$\pfourffour$ starts overlapping with $\pgrcgr$. So for high $\gamma$
flows, along with $\Phi_1(r)$, $\Phi_4(r)$ may also be considered as a
plausible approximation for general relativistic accretion. 
Shocked flows in $\Phi_2(r)$ and $\Phi_3(r)$ {\it never} show any
overlap with $\pgrcgr$ for any value of $\gamma$; hence these 
potentials may not be considered to mimick the  general
relativistic multi-transonic accretion flows.
\section{Concluding Remarks}
\noindent
In this paper, we provide a generalized formalism which
can formulate and solve the equations governing the
advective, multi-transonic, hydrodynamic BH accretion
in {\it all} available pseudo-Schwarzschild potentials, which
may contain steady, standing, Rankine-Hugoniot kind of shocks.
We have also formulated and solved the equations governing
multi-transonic, complete general relativistic BH accretion
and wind in Schwarzschild metric and compared our pseudo-Schwarzschild 
solutions with the general relativistic one. The main conclusions of this
paper are the following:\\
\noindent
(a) We
observe that a significant region of
parameter space 
(spanned by the conserved total specific energy ${\cal E}$,
the specific angular momentum $\lambda$ and the polytropic
index $\gamma$ of the flow) allows shock formation for {\it all}
potentials, which leads to the strong conclusion that
stable, standing RHS are
inevitable ingredients in multi-transonic
accretion disks around non-rotating
BHs. The same kind of conclusion was drawn by previous works in 
this field  (see \S 1) {\it only} for 
ultra-relativistic accretion in Paczy\'nski \& Wiita 
(1980) potential, whereas we make this conclusion more general
by incorporating {\it all} available BH potentials to
study BH  accretion  for {\it all}  possible values of
$\gamma$.\\
\noindent
(b) As the shock forms at a particular radial
distance,
it is clear that self-similar solutions
should {\it not} be invoked while studying real physical BH
accretion and related phenomena.\\
\noindent
(c) It is sometimes
argued that a non-standing oscillating shock may
modulate the disc spectrum in order to explain the
dwarf novae outburst (Mausche, Raymond \& Mattei 1995)
or QPO (Hua, Kazanas \& Titarchuk 1997).
In this context, the region of parameter space, for which 
three sonic points are formed in accretion but 
still 
no steady, standing shock is found (see \S 3), 
can be considered as  quite an important zone because
$\left[{\cal P}_i\right]\in
{\cal G}_i\left[{\cal P}_i\right]$ may provide the relevant parameters
responsible for such physical processes.\\
\noindent
(d) As long as the shock formation in ultra-relativistic
black hole accretion is concerned, 
the Paczy\'nski \& Wiita (1980) potential $\Phi_1(r)$
is the {\it only} pseudo potential which can mimic the solutions of
general relativistic accretion disc around non-rotating BHs in a very
efficient way. However, in the purely non-relativistic limit
($\gamma{\longrightarrow}5/3$), along with $\Phi_1(r)$, another 
BH potential $\Phi_4(r)$ proposed by ABN, is also observed to
mimic the general relativistic solutions;
at least for low energy - high angular momentum flows. However,
it is interesting to note one important feature
of the Paczy\'nski \& Wiita potential $\Phi_1(r)$; like spherically 
symmetric accretion (see DS), for 
accretion disc also, $\Phi_1(r)$ is observed to be in excellent 
agrement with solutions for ultra-relativistic 
flow in pure Schwarzschild metric, however, it starts
loosing (albeight very slowly) its efficiency in mimicking 
full general relativistic solution with higher values of $\gamma$, 
i.e., as the flow reaches its purely non-relativistic limits;
although the exact reason behind this is not quite clear to us.\\
\noindent
Hot, dense and
exo-entropic post-shock regions in advective accretion
disks are used as a powerful tool in understanding the
spectral properties of BH candidates (Shrader \& Titarchuk 1998, and
references therein)
and in theoretically explaining a
number of diverse phenomena, including millisecond
variability in the X-ray emission from LMXBs and the
generation mechanism for high frequency QPOs in
general (Titarchuk, Lapidus \& Muslimov 1998 and
references therein), high energy emission from central
engines of AGNs (Sivron, Caditz \& Tsuruta 1996),
formation of heavier elements in BH accretion discs
via non-explosive nucleosynthesis (Mukhopadhyay \&
Chakrabarti 2000), formation and dynamics of accretion
powered galactic and extragalactic jets, quiescent
states of X-ray novae systems and outflow induced low
luminosity of our galactic centre (Das 2001; Das
\& Chakrabarti 1999). A number of
observational evidences are also present which are in
close agreement with the theoretical predictions
obtained from shocked accretion model (Rutledge et al.
1999; Muno, Morgan \& Remillard 1999;
Webb \& Malkan 2000; Rao, Yadav \&
Paul 2000; Smith, Heindl \& Swank 2001).
Thus we believe that our present work may
have far reaching consequences because 
of the following reasons:
\begin{enumerate}
\item Our generalized formalism assures that
that our model is
not just an artifact of a particular type of potential only
and inclusion of every BH potential allows
a substantially extended zone of parameter space
allowing for the
possibility of shock formation.
\item Of course there are possibilities that in future someone
may come up with a pseudo-Schwarzschild potential better than $\Phi_1(r)$,
which will be the best approximation for complete general relativistic
investigation of multi-transonic shocked flow. In such case, if one 
already formulates a generalized model for multi-transonic shocked 
accretion disc
for any arbitrary $\Phi(r)$, exactly what we have done in this paper,
then that generalized model will be able to readily accommodate that new
$\Phi(r)$ without having any significant change in the fundamental
structure of the formulation and solution scheme of the model
and we need not
have to worry about providing any new scheme exclusively valid only for
that new potential, if any.
\item Even if someone can provide a completely satisfactory
model for shock formation in full general relativistic (Schwarzschild) BH 
accretion, still the utility of this work may not be
completely irrelevant. Rigorous investigation of some  of the 
shock related phenomena is extremely difficult (if not completely
impossible) to study using full general relativistic framework. Hence one 
is expected to always rely on these pseudo-potentials because of the 
ease of handling them. For example, it was shown that
(see \S 4) the total energy of the general relativistic accretion 
flow can {\it not} be decoupled into its constituent contributions,
whereas for any kind of pseudo-potential (see \S 2), all 
individual energy components are linear under addition. This 
provides enough freedom and ease to simply add any extra 
component in the expression for energy to introduce any new 
physics in the system (radiative forces or magnetic fields for example),
which is certainly not possible while dealing with full general relativistic
astrophysical flows around non-rotating BHs. 
\end{enumerate}
Thus, for above mentioned reasons, we believe that 
compared to all previous works based solely on ultra-relativistic accretion
in
$\Phi_1$, our model  is
better equipped for handling various
shock related phenomena.\\
\noindent
It is noteworthy that the idea of shock formation in advective BH accretion
is contested by some authors (Narayan, Kato \& Honma 1997, and references
therein). However, the fact that their claim against shock formation
is,  perhaps, inappropriate for many reasons, has been shown
(Molteni, Gerardi \& Valenza 2001) from energy considerations. One can 
understand that the problem of not finding shocks lies in the fact that 
non-shock ADAF models are, perhaps, unable to produce multi-transonic
flows because only one inner sonic point close to the BH is explored
by such works. \\
\noindent
One can observe that flows characterized by 
$\left[{\cal P}_i\right]\in
{\cal F}_i\left[{\cal P}_i\right]$ in our work may contain low intrinsic
angular momentum for some cases (especially for purely
non-relativistic flow in some of the BH potentials)
However, such weakly rotating flows are expected to be allowed by nature 
for various real physical situations like detached binary systems
fed by accretion from OB stellar winds (Illarionov \& 
Sunyaev 1975; Liang \& Nolan 1984), semi-detached low-mass
non-magnetic binaries (Bisikalo et al.\ 1998) and supermassive BHs fed
by accretion from slowly rotating central stellar clusters
(Illarionov 1988; Ho 1999 and references therein). \\
\noindent
Even 
twenty-eight years after the discovery of standard accretion disc theory
(Shakura \& Sunyaev 1973), exact modeling of viscous multi-transonic
BH accretion, including proper heating and cooling mechanisms is still 
quite an arduous task, and we have not yet fully attempted this.
However, our preliminary calculations show that the introduction of viscosity
via a radius dependent power law distribution for angular momentum
pushes the shock location closer to the BH; details of this work will be
discussed elsewhere.

\acknowledgments
\noindent
The author acknowledges discussions with
Igor D. Novikov, A. R. Rao and Sandip K. Chakrabarti.
His very special thanks goes to Paul J. Wiita for reading the
manuscript extremely carefully and for providing a number of 
useful suggesstions. He is also thankful to 
Aveek Sarkar for checking some of the algebra. Finally, the
hospitality provided by the Racah Institute of Physics, The 
Hebrew University of Jerusalem, Israel, is acknowledged, where
a part of this paper was written.

{}
\newpage
\begin{center}
{\large\bf Figure Captions} \\[1cm]
\end{center}
\noindent 
Fig.\ 1:  Effective BH potentials for general relativistic
($\Phi_{BH}^{eff}(r)$) as well as for pseudo-general
relativistic($\Phi^{eff}_i(r)$)
 accretion discs as a function of the distance
(measured from the event horizon in units or $r_g$) plotted 
in logarithmic scale. The specific angular momentum  is   chosen to 
be 2 in geometric units.  See text for details.

\noindent
Fig.\  2: Parameter space division for multi-transonic, 
ultra-relativistic accretion and wind in four different 
pseudo-Schwarzschild BH potentials, see text for details.

\noindent Fig.\ 3: Solution topologies for multi-transonic,
ultra-relativistic ($\gamma=4/3$)
shocked flows in different BH
potentials as indicated in the figure.  See text for details.

\noindent
 Fig.\ 4:  Region of parameter space responsible for shock 
formation (${\cal F}_i\left[{\cal P}_i\right]$), for four different 
BH potentials $\Phi_1$ (a), $\Phi_2$ (b), $\Phi_3$ (c), and
$\Phi_4$ (d).  See text for details.

\noindent
Fig.\ 5: Parameter space division for ultra-relativistic,
multi-transonic, accretion and wind in general relativity.

\noindent
Fig.\ 6: Integral curves of motion for ultra-relativistic, 
multi-transonic, black hole accretion and corresponding `self-wind' in 
Schwarzschild metric. See text for details.

\noindent
Fig.\ 7: Comparison of parameter space producing shocked multi-transonic
accretion in  various BH potentials, with parameter space representing 
multi-transonic black hole  accretion  and wind in general relativity.
The figure is drawn for the ultra-relativistic flow,  see 
text for details.
\newpage
\plotone{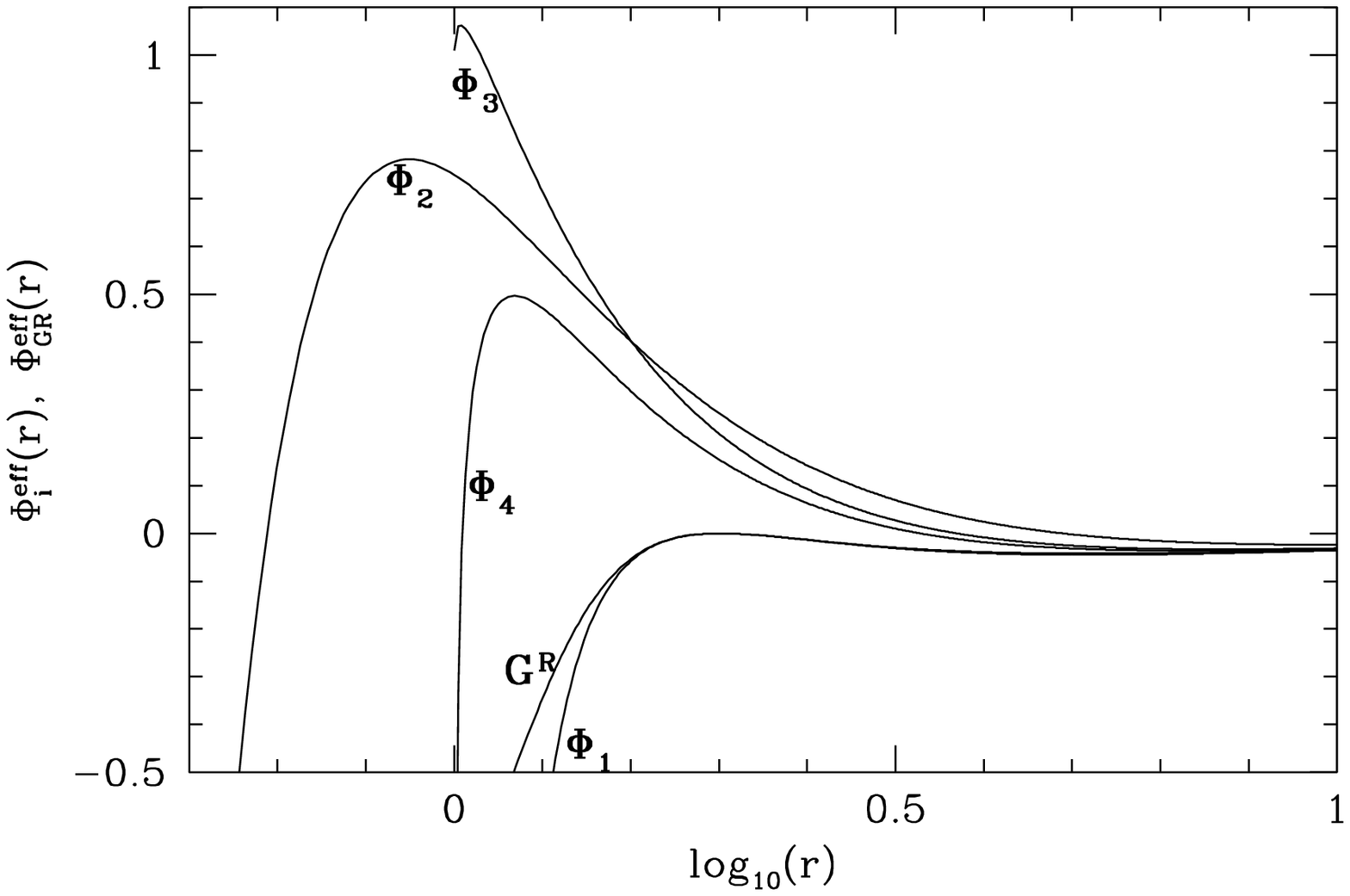}

\newpage
\plotone{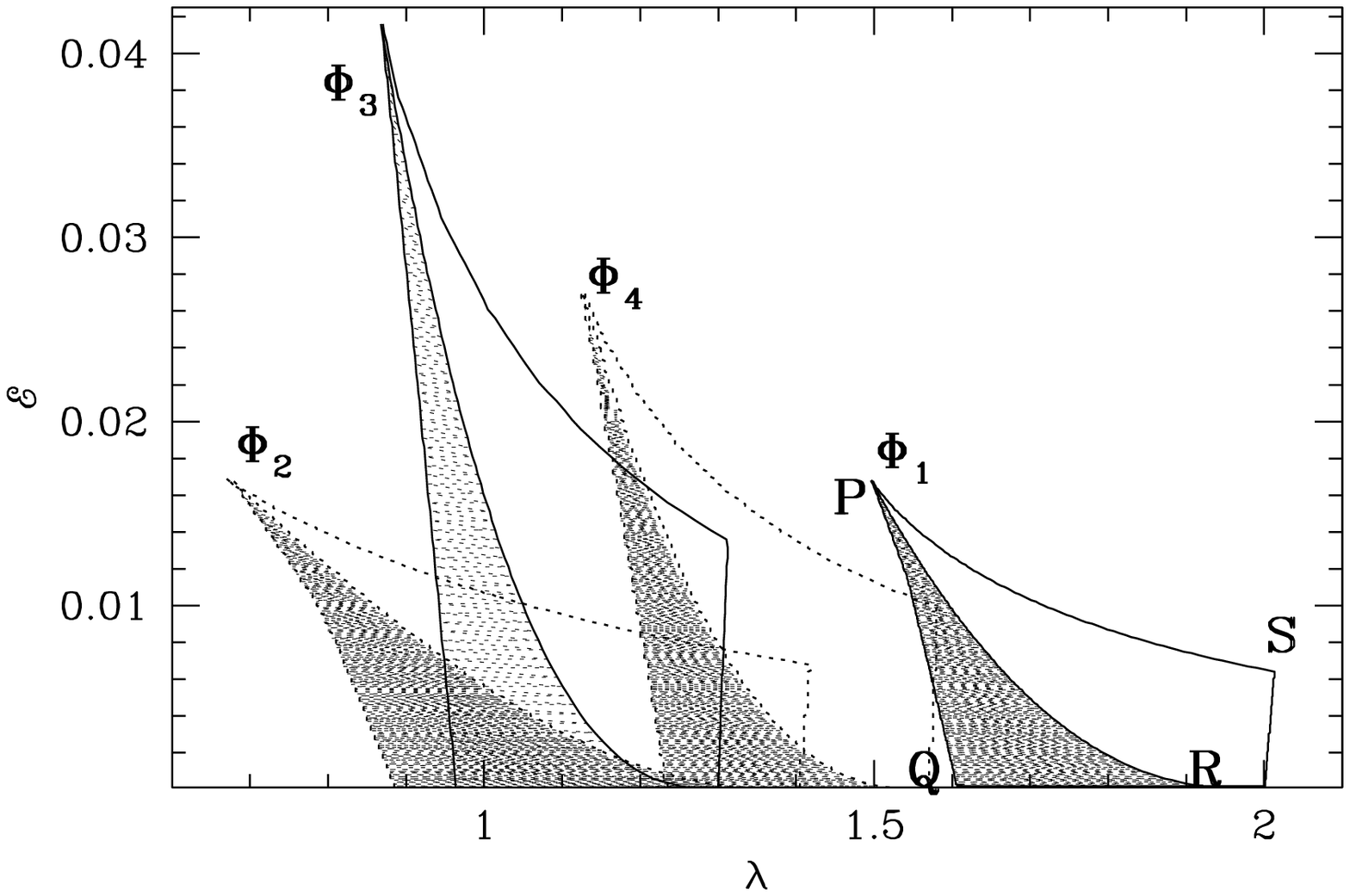}

\newpage
\plotone{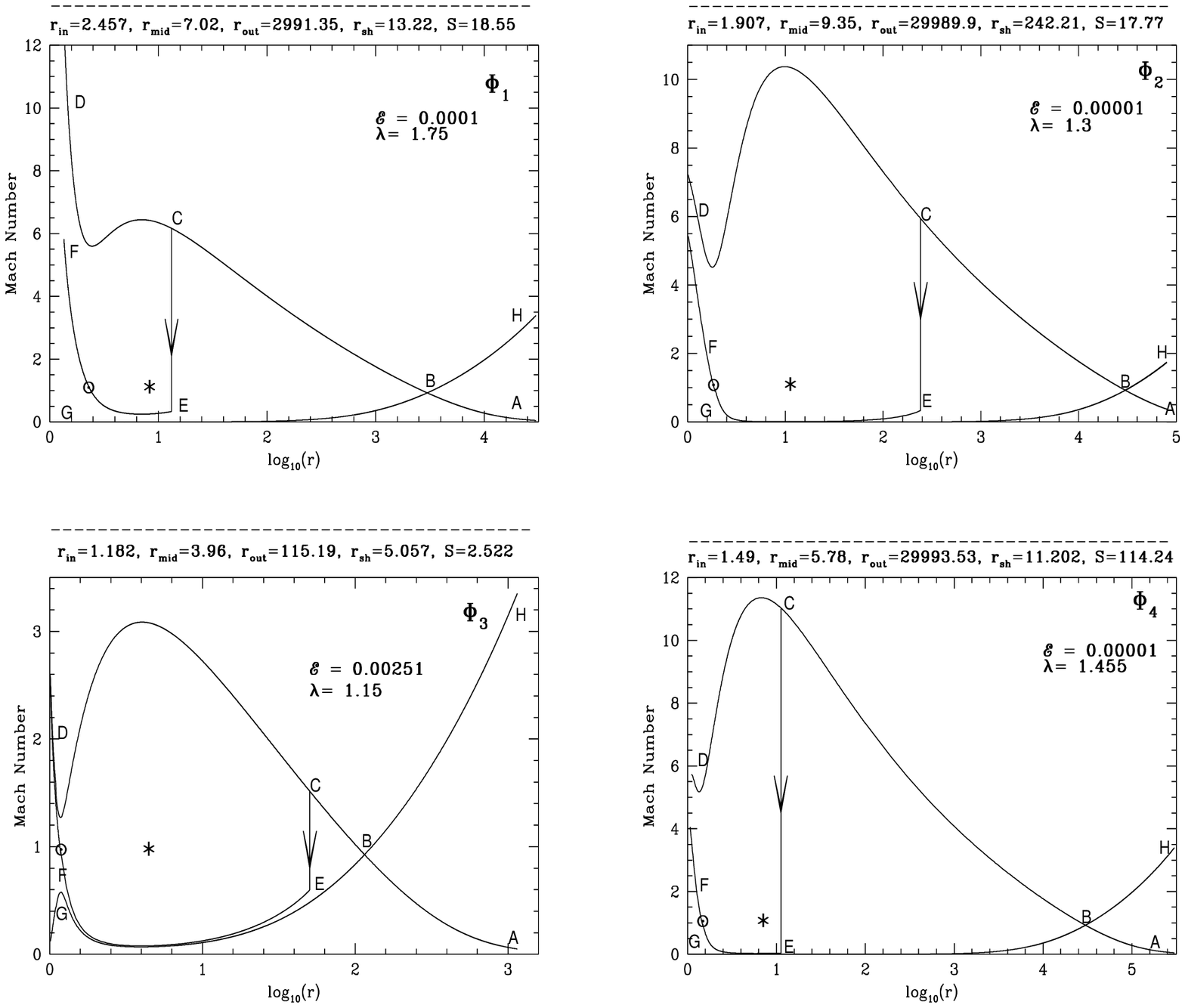}

\newpage
\plotone{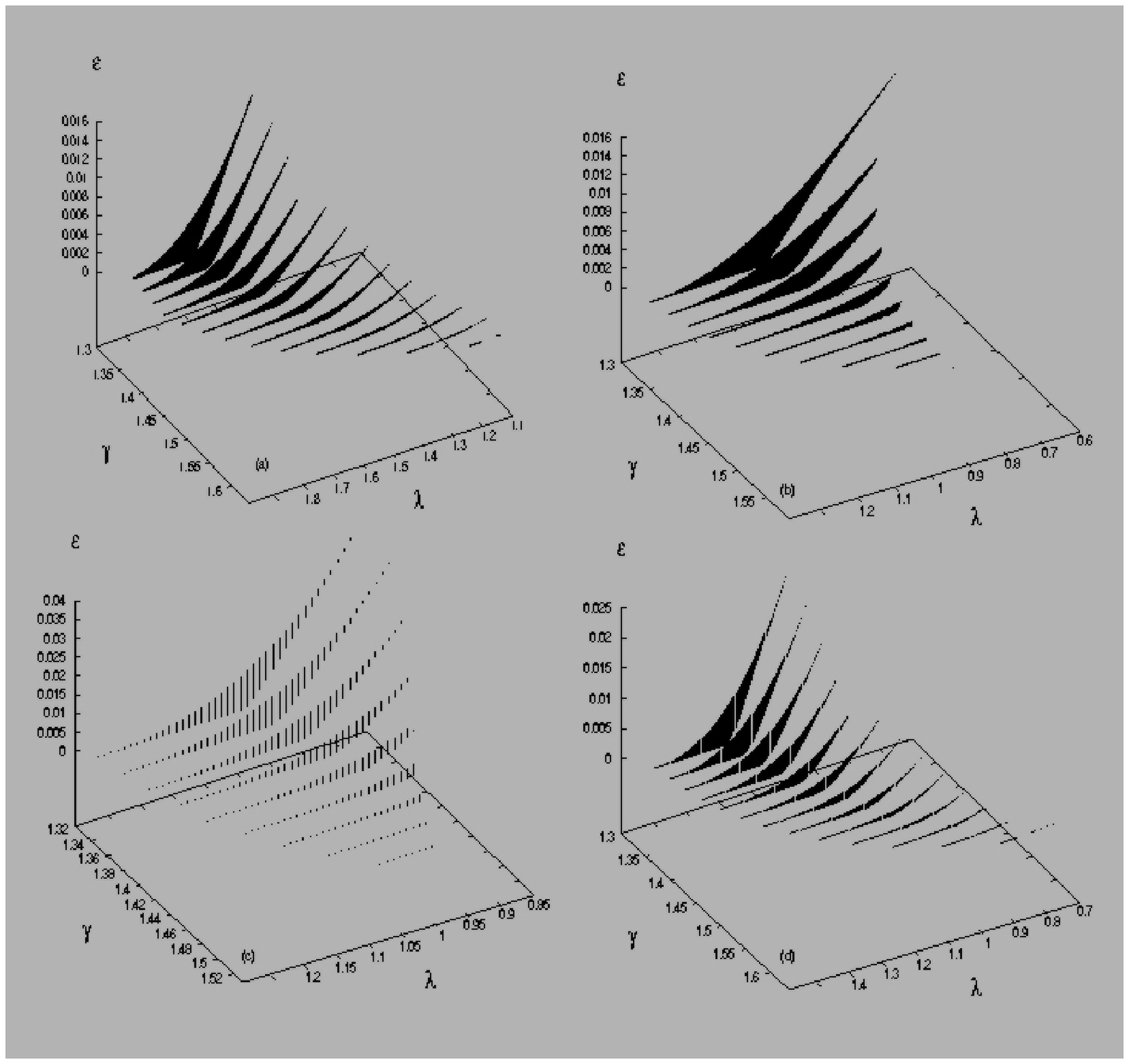}

\newpage
\plotone{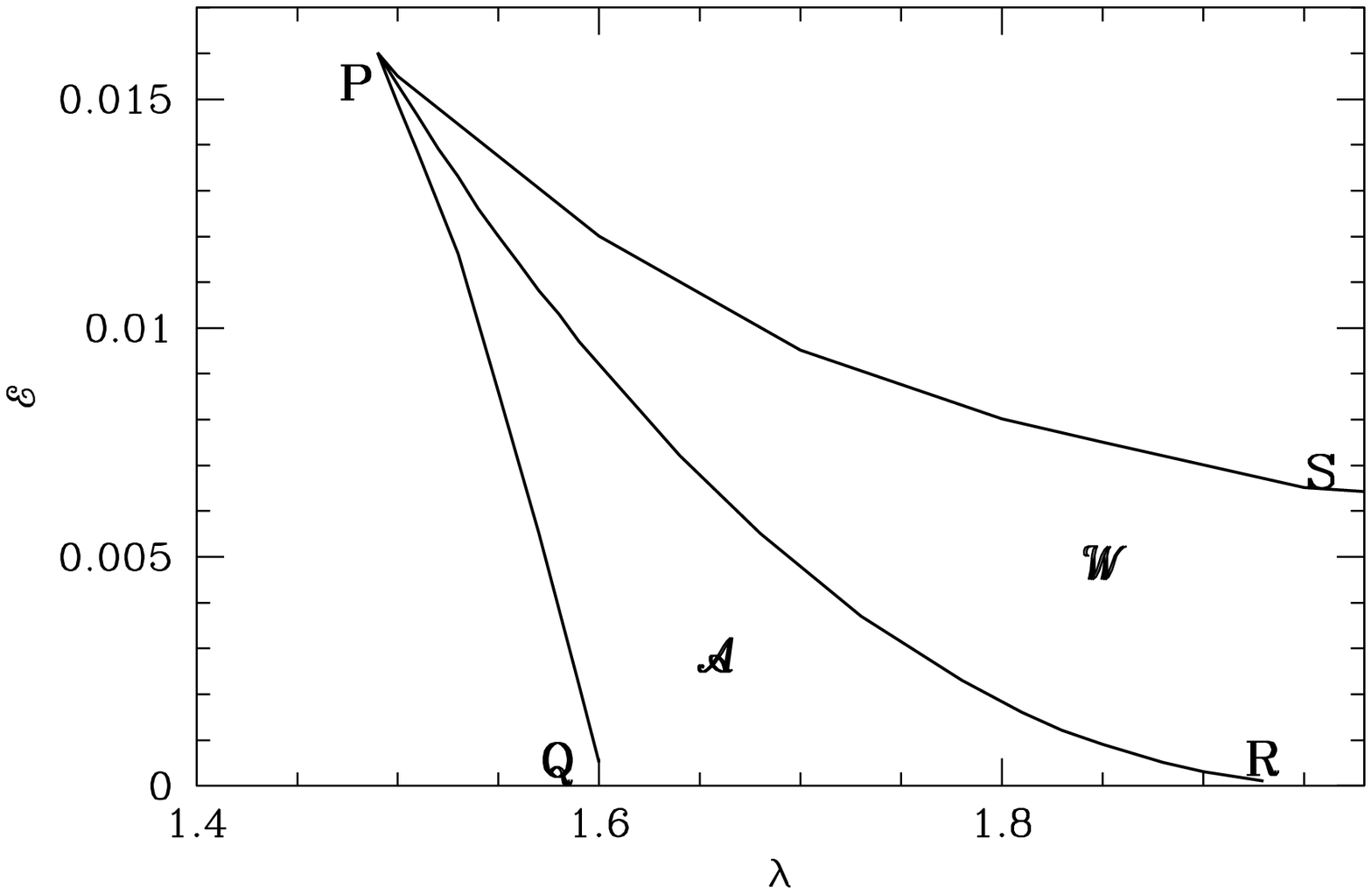}

\newpage
\plotone{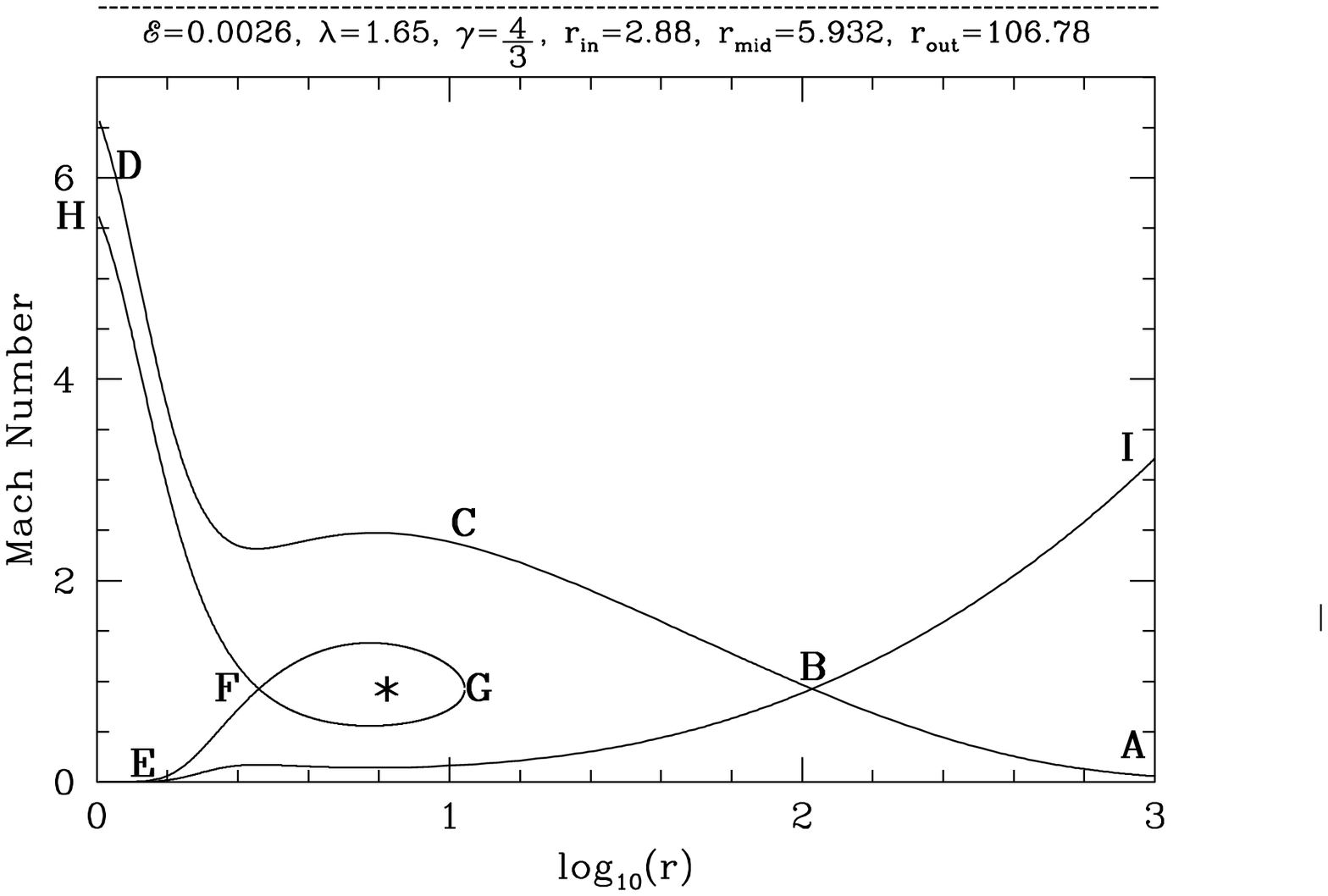}

\newpage
\plotone{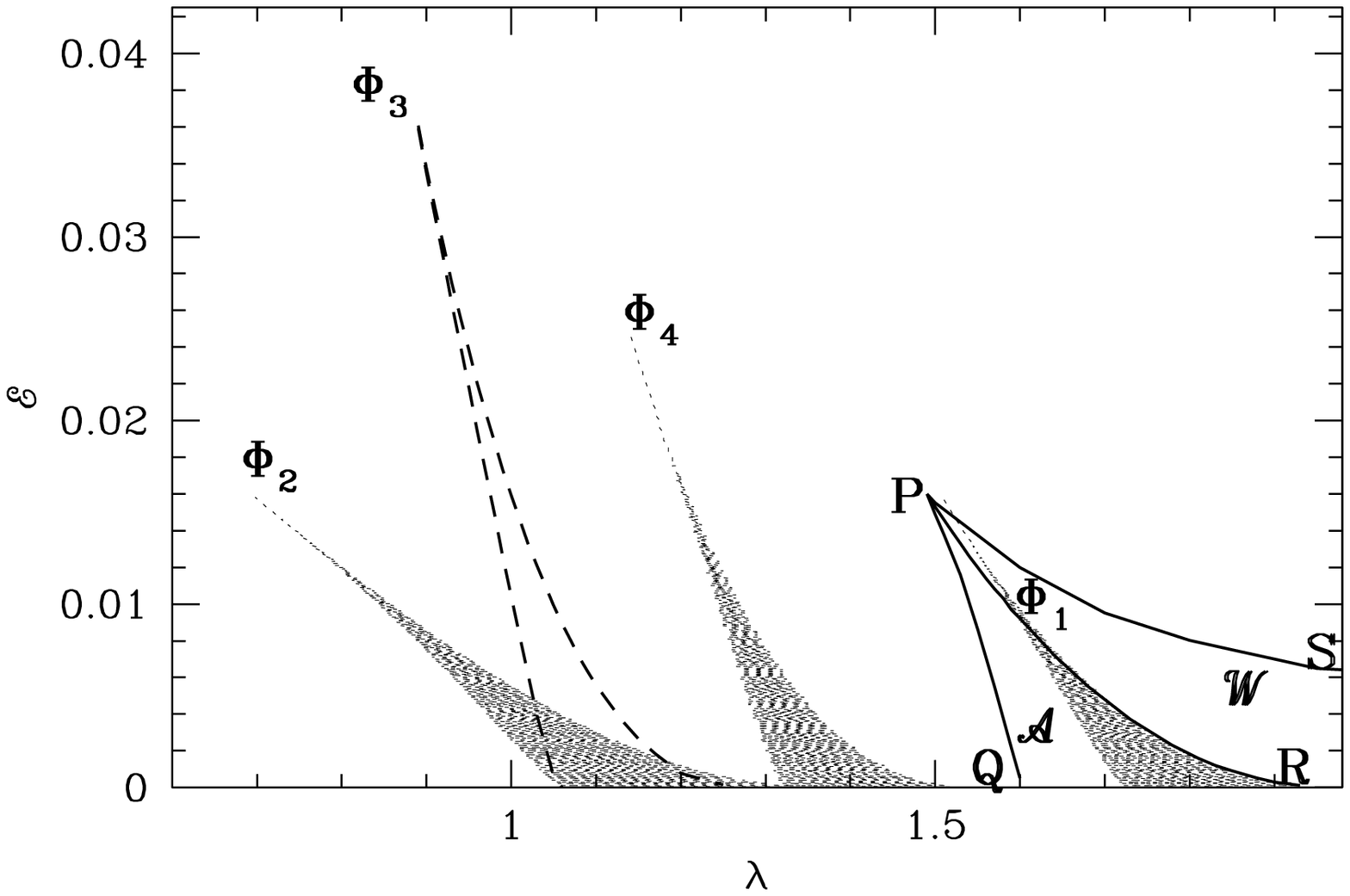}

\end{document}